\documentclass[aps,prl,twocolumn,superscriptaddress,showpacs]{revtex4}

\newcommand{\w}{{\omega}}
\newcommand{\s}{{\sigma}}

\def\be{\begin{eqnarray}}
\def\ee{\end{eqnarray}}
\newcommand{\nn}{\nonumber\\}

\newcommand{\Eq}[1]{Eq.~(\ref{#1})}

\newcommand{\ra}{\rightarrow}

\newcommand{\ket}[1]{|#1\rangle}

\usepackage{graphicx}

\begin{document}

\title{Photoisomerization in a Dissipative Environment}

\author{Chang-Qin Wu}
\affiliation{Department of Physics, University of California at
Berkeley, Berkeley, CA 94720, USA} \affiliation{Department of
Physics, Fudan University, Shanghai 200433, China}
\author{Jian-Xin Li}
\affiliation{Department of Physics, University of California at
Berkeley, Berkeley, CA 94720, USA} \affiliation{National Laboratory
of Solid State Microstructure, Nanjing University, Nanjing 210093,
China}
\author{Dung-Hai Lee}
\affiliation{Department of Physics, University of California at
Berkeley, Berkeley, CA 94720, USA} \affiliation{Material Science
Division, Lawrence Berkeley National Laboratory, Berkeley, CA 94720,
USA}

\date{\today}

\begin{abstract}
We investigate photoisomerization (PI), the shape change of a
molecule upon photoabsorption, in a dissipative environment using
a simple spin-boson model. We identify two classes of environment
depending on whether it "entangles" with the molecule. In the
absence of entanglement, the environment merely causes a blue
shift of the required photon frequency and reduces the quantum
efficiency of PI. With entanglement the molecule can undergo a
quantum phase transition between a state that photoisomerizes to a
state that does not.
\end{abstract}



\maketitle

Photoisomerization (PI) is a phenomenon where a molecule absorbs a
photon and changes its shape
(conformation).\cite{rau,jiang,hasson,vogt,hoki} This important
phenomenon is relevant to vision and other biological processes.
Recently it has been proposed that molecules which photoisomerize
(the "PI molecules") can be used as light sensitive molecular
switches\cite{zhang,choi} and optomechanical devices.~\cite{hugel}
A good example of PI molecule is ``azobenzene''.\cite{rau} It
switches between two isomers, the \emph{trans} and \emph{cis}
configurations upon absorbing light. 

For microscopic systems, environment usually plays a critical role
in affecting their quantum mechanical behavior. PI of molecules is
not an exception. The goal of this paper is to study PI in the
presence of an environment. Although the rest of the discussions are
coined in terms of PI, our theory should be applicable to a variety
of situations. Examples include Jahn-Teller distortion in an
environment and quantum bit in an environment.

The physical mechanism of PI has attracted a lot of interests in the
past two decades. In this work we present a simple theory for it.
Our theory differs from those in the literature in that we emphasize
the importance of the environment. 
Specifically, we suggest the existence of two types of environment,
depending on whether it ``entangles'' with the PI molecule. Without
entanglement the environment merely causes a blue shift in the
photon frequency necessary to induce PI, and reduces the quantum
efficiency. With
entanglement the molecule can undergo a quantum phase transition 
between a state that photoisomerizes to a state that does not. Our
investigation of the environmental effects on PI is stimulated by
a very recent scan tunneling microscopy experiment\cite{crommie}
which reported the loss of PI functionality when azobenzene is in
close contact with a metallic (gold) surface, and the regaining of
the PI functionality as the molecules are elevated sufficiently
far from the surface.


The irreducible ingredients of a PI molecule are (1) two
electronic states\cite{note} and (2) a structure degree of
freedom, and (3) a strong coupling between the two. We model (1)
by a quantum bit (or a quantum spin) where the up and down states
correspond to the two electronic states in question, and (2) by a
simple harmonic oscillator. The coupling between the two sets of
freedom is described by the following Hamiltonian \be H_{\rm
mol}&&=t_0\sigma_z+g' \sigma_z x+
{k\over 2}x^2+{p^2\over 2m}.
\label{hm} \ee Here $x$ and $p$ are the dynamic variables of the
harmonic oscillator, $\s_z$ is the third Pauli
matrix, and 
$2t_0$ describes the energy difference between the two electronic
states. It is instructive to first analyze \Eq{hm} in the limit
$m\rightarrow\infty$ where the harmonic oscillator is classical. In
this limit, depending on the value of $\s_z$ ($\pm 1$), the energy
landscape for the structural degree of freedom ($U(x)$) takes on
different shapes. This is shown as the red and green curves in
Fig.~1(a).
In conventional picture of PI, the system jumps vertically from the
minimum of the red (green) curve to the green (red) curve upon
absorbing a photon. Subsequently the structure relaxes until $x$
reaches the minimum of the green (red) curve. The two minima of the
green and red curves, situated at $x=\pm g'/k$, correspond to the
two relevant molecular structures. We note that by setting $t_0$ to
zero \Eq{hm} describes the physics of Jahn-Teller effect where the
structure distortion (i.e., $x=0\ra x\ne 0$) spontaneously sets in
to split the electronic degeneracy and lower the total energy.

The above intuitive picture for PI leaves unclear an important
question: since electronic transition often proceeds at a faster
rate than structural relaxation, what holds the molecule from
de-exciting back to the original electronic state during structural
relaxation? We begin our discussions by first addressing this
question.

Restoring the full quantum mechanical nature of the harmonic
oscillator and introducing creation ($a^\dagger$) and annihilation
($a$) operators, we replace the 2nd to 4th terms of \Eq{hm} by $g_0
\sigma_z (\hat{a}^\dagger+\hat{a})+\omega_0 \hat{a}^\dagger\hat{a}.$
This simple model can be solved exactly to yield the following
eigen-states and eigen-energies \be
&&|s,n\rangle=\frac{e^{-a_c^2/2}}{\sqrt{n!}}
(\hat{a}^\dagger+sa_c)^ne^{-sa_c\hat{a}^\dagger}|s,0\rangle,\nn
&&E_{s,n}=n\omega_0+st_0-g_0^2/\omega_0. \label{eig}\ee Here
$a_c\equiv g_0/\omega_0$, $s=\pm 1$ (which labels the two electronic
states), and $n$ are non-negative integers. The state $\ket{s,0}$
satisfies $\hat{a}|s,0\rangle=0$ and
$\sigma_z|s,0\rangle=s|s,0\rangle$. It is simple to show that
regardless of the value for $n$, we have $\langle \pm 1,n |x|\pm 1,n
\rangle=\mp 2 a_c$ implying the presence of two structural
conformations with the associated quantum structural fluctuations.
For $t_0>0$, the ground state is $|-1,0\rangle$ with $\langle x
\rangle=2a_c$. After absorbing a photon the molecule make a
transition from $|-1,0\rangle$ to $\ket{+1,n}$. When that happens
the structure changes from that corresponding to $\langle
x\rangle=+2a_c$ to that corresponding to
$\langle x\rangle=-2 a_c$. 

If we model the electric dipole operator as $\s_x$, the optical
absorption spectrum is given by  \be
\sigma(\varepsilon)=\sum_{n=0}^{\infty} |\langle
+1,n|\sigma_x|-1,0\rangle|^2\delta(\varepsilon-E_{+1,n}+E_{-1,0}).
\label{abs} \ee
Substituting \Eq{eig} into \Eq{abs} we obtain $
\sigma(\varepsilon)=\sum_{n=0}^\infty
f_n(a_c)\delta(\varepsilon-n\omega_0-2t_0), $ where
$f_n(a_c)=(4a_c^2)^n e^{-4a_c^2}/n!$ is the Franck-Condon factor
between the initial $\ket{-1,0}$ and final $\ket{+1,n}$ states. The
red curve in Fig.~1(b) shows $\sigma(\epsilon)$ for a specific
choice of parameters. The closely spaced peaks originate from the
excited states of the harmonic oscillator.
\begin{figure}
\includegraphics[angle=0,scale=1]{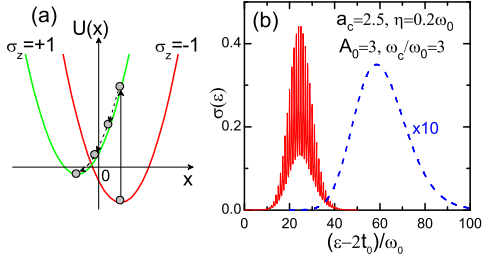}
\caption{(Color online) (a)The energy landscape for the structure
coordinate.(b) The optical absorption spectrum
$\sigma(\varepsilon)$ for the PI molecule without (red line) and
with (blue line) coupling to environments. $\eta$ is the Lorentz
broadening parameter.\label{fig1}}
\end{figure}
The envelope of the red absorption curve is determined by the
Franck-Condon factor $f_n(a_c)$. As shown in
Fig.~1(b) the absorption practically vanishes for
$(\varepsilon-2t_0)/\omega_0\lesssim 10$. This is due to the
suppression of the optical transition matrix element by $f_n(a_c)$.
Likewise, the envelope peaks at the energy where the Frank-Condon
factor reaches maximum. At this peak frequency the incident photon
has the highest quantum efficiency in inducing isomerization. Due to
the strong suppression of the optical matrix element, the molecule
will cease to spontaneous emit once its energy has descended below
certain energy ($\sim 10\omega_0$ in Fig.~1(b)). This result
suggests the presence of two time scales
$T_{1,2}$. 
$T_1$ is the time it takes for {\it structural relaxation} to
bring the molecule's energy outside the absorption peak, and $T_2$
is the time for it to bring the molecule to the isomerized ground
state. During $t=0\ra T_1$ the molecule relaxes both structurally
and electronically (by spontaneous emission). During $t=T_1\ra
T_2$ the molecule can only relax structurally because spontaneous
emission is strong suppressed. Thus once the photo-excited
molecule survives through $T_1$ it can relax to the isomerized
ground state without the further interruption from spontaneous
emission. This answers the earlier posed question, and at the same
time predicts that the quantum efficiency of PI is
determined by the probability for the molecule to survive through $T_1$.
The existence of time regimes with different relaxation dynamics is
consistent with the experimental results for
azobenzene.\cite{nagele} Moreover, due to the energy difference
$2t_0$, the peak photon frequency for the $\langle x\rangle=+2
a_c\rightarrow\langle x\rangle=-2 a_c$ conversion will be different
from the $\langle x\rangle=-2 a_c\rightarrow\langle x\rangle=+2 a_c$
one. This is also consistent with experiments.\cite{nagele}


Next we consider the effects of the environment. Following Calderia
and Leggett\cite{CL} we model the environment by a continuum of
harmonic oscillators. These oscillators couple to the PI molecule
via the Hamiltonian
\begin{equation}
H_{\rm env}=\left(\sum_\alpha g_\alpha
(a^\dagger_\alpha+a_\alpha)\right)\sigma_\mu+\sum_\alpha\w_\alpha
a^\dagger_\alpha a_\alpha.\label{cp}
\end{equation}
Here $\alpha$ labels the environment oscillators and $\mu=$ x or y
or z depending on the nature of molecule-environment coupling.
Clearly when $\mu=z$ we have $[H_{\rm mol},H_{\rm env}]=0$, which
implies the absence of ``entanglement'' between the molecule and the
environment (see later). On the other hand, for $\mu= x$ or $y$
$H_{\rm mol}$ and $H_{\rm env}$ do not commute and the environment
entangles with the molecule. As we shall show that these two types
of environment have fundamentally different effects on PI.

Let us begin with the unentangled environment (i.e. $\mu=z$).
In this case the eigenstates of the molecule plus the environment
are the following direct products (hence no entanglement) $
\ket{s,n}\otimes\ket{s,\{n_\alpha\}}.$ Here $\ket{s,n}$ is given in
\Eq{eig} and $\ket{s,\{n_\alpha\}}$ is given by \be
\ket{s,\{n_\alpha\}}=\prod_\alpha\frac{e^{-{a_c^\alpha}^2/2}}{\sqrt{n_\alpha!}}
(\hat{a}^\dagger_\alpha+sa_c^\alpha)^ne^{-sa_c^\alpha\hat{a}^\dagger_\alpha}|s,\{0\}\rangle\ee
with $a_c^\alpha\equiv g_\alpha/\omega_\alpha$.  In this case the
environment has no effect on the structure, i.e., $\langle
x\rangle$, of the molecule. It merely modifies the absorption
spectrum to
\begin{equation}
\sigma(\varepsilon)=\sum_{n=0}^\infty
f_n(a_c)D(\varepsilon-n\omega_0-2t_0),
\end{equation}
where $D(\omega)=\int_{-\infty}^\infty {dt\over 2\pi} e^{-i\omega
t+4K(t)},~~K(t)=\int_0^\infty
\frac{d\omega}{\omega^2}J(\omega)(e^{i\omega t}-1),$ and
$J(\omega)=\sum_\alpha g_\alpha^2\delta(\omega-\omega_\alpha).$ In
the case where the environmental dissipation is Ohmic\cite{leggett}
we have $ J(\omega)=A_0\omega e^{-\omega/\omega_c}$, where $A_0$
reflects the molecule-environment coupling strength, and $\omega_c$
is a cut-off frequency. For, e.g, $A_0=3$ and $\w_c/\w_0=3$ the
absorption spectrum is shown as the blue curve in Fig.~1(b). It is
blue-shifted relative to the red curve, and the peak intensity is
greatly reduced reflecting a considerable reduction of the quantum
efficiency of PI. Moreover, the sharp phonon peaks in the red curve
is now smoothened.

Next, we consider the more interesting case where the environment
entangles with the molecule. Such an environment can be simulated by
putting $\mu=x$ in \Eq{cp}, which results in the non-commutivity
between $H_{\rm mol}$ and $H_{\rm env}$. Physically this type of
environment induces transition between the electronic states (or
flips the spin). When resent alone, the ground state of $H_{\rm
mol}$ and $H_{\rm env}$ is the eigenstate of $\s_z$ and $\s_x$
respectively. When present together the spin is subjected to the
conflicting quantum noise exerted by the structure degree of freedom
 and the environment. If we choose the eigenstate of, say,
$\s_x$ to describe the spin, the dynamics dictated by $H_{\rm
mol}+H_{\rm env}$ causes the spin to flip as a function of time. The
primary effect of the environment is to induce correlation between
spin-flipping events. In the following we use a well-established
formulation\cite{leggett} which translates above physical picture
into precise mathematical languages. Technically we employ the
imaginary-time path integral representation of the quantum partition
function of the molecule plus the environment. We then ``integrate
out'' the environmental degrees of freedom and obtain the following
effective partition function for the molecule
\begin{widetext}
\begin{equation}
Z=\sum_{n=0}^\infty \int_0^\beta d\tau_{2n}
\int_0^{\tau_{2n}-\tau_{0}} d\tau_{2n-1}\cdots
\int_0^{\tau_{2}-\tau_0} d\tau_{1}
\int\mathcal{D}[x(\tau)]\prod_{i=1}^{2n}[t_0+g'x(\tau_i)]~e^{-S[x(\tau),Q(\tau_i)]},
\label{z}
\end{equation}
\end{widetext}
here $2n$ counts the number of $\s_x\ra -\s_x$ spin flips between
$\tau=0$ and $\tau=\beta$, the imaginary time integral is subjected
to the contraint $\tau_1<\tau_2<\cdots<\tau_{2n}$, $x(\tau)$
specifies the history of the molecular oscillator, and
$\tau_0=1/\omega_c$.
The quantity $Q(\tau_i)$ is an alternating sequence of $\pm 1$ ($+1$
for down $\ra$ up flip and $-1$ for up $\ra$ down flip respectively)
at $t_1,..,t_{2n}$. The action $S[x(\tau),Q(\tau_i)]$ in \Eq{z} is
given by
\begin{equation}
S=\int_0^\beta d\tau H(\tau)-\sum_{i\ne j}Q(\tau_i)
V(\tau_i-\tau_j)Q(\tau_j), \label{action}
\end{equation}
where $V(\tau)=\int_0^\infty d\omega \omega^{-2}
J(\omega)\exp(-\omega|\tau|)$ and $H(\tau)={m\over
2}[\dot{x}(\tau)^2+\w_0^2 x(\tau)^2]$. 
In the case of the Ohmic environment ($ J(\omega)=A_0\omega
e^{-\omega/\omega_c}$) we obtain
\begin{equation}
V(\tau_{i}-\tau_{j})=A_0 \ln
\left(\frac{|\tau_{i}-\tau_{j}|+\tau_0}{\tau_0}\right).\label{log}
\end{equation}
In the limit where $t_0\tau_0$ and $g_0\tau_0$ are small, a
well-known renormalization group technique\cite{anderson} has been
developed to treat \Eq{z}. In this treatment one progressively
integrates out pairs of up and down spin flips that are separated
farther and farther in time, and record the effects of such
integration on the parameters of \Eq{action}. For $V(\tau_i-\tau_j)$
given by \Eq{log} the result of such calculation gives the following
recursion relations $\frac{dA}{dl}=-2A(t^{2}+g^{2}e^{-\omega}),~
\frac{dt}{dl}=(1-A/2)t,~ \frac{dg}{dl}=(1-A/2)g, ~
\frac{d\omega}{dl}=\omega$, here $l$ is the logarithm of the time
separation of the farthest spin flip pair integrated out so far
divided by $\tau_0$, and $A,~t,~g$, and $\omega$ are all
dimensionless quantities. They are function of $l$ satisfying the
initial conditions
$A(0)=A_0,~t(0)=t_0\tau_0,~g(0)=g_0\tau_0,~\omega(0)=\omega_0\tau_0$.
These recursion relations can be solved numerically to obtain the
phase diagram for the zero temperature ($\beta\ra\infty$) state of
the PI molecule as a function of the initial $t_0\tau_0$ and $A_0$
for different $g_0/t_0$ and $\w_0\tau_0$. In Fig.~2 we present the
phase boundary between the ``instanton metal'' where the spin flips
are uncorrelated (aside from the global constraint of alternating up
and down spin flips), and the ``instanton insulator'' where up and
down spin flips are correlated so that they form bound pairs in
time, for several different values of $\w_0\tau_0$ at a fixed
$g_0/t_0$.

To obtain information about the molecular structure in these two
phases we compute the correlation function
$\Gamma_{xx}(\tau_a-\tau_b)=\langle x(\tau_a)x(\tau_b)\rangle$. It
can be shown that \be
&&\Gamma_{xx}(\tau_a-\tau_b)=G_0(\tau_a-\tau_b)\nn&&+g'^2\int
d\tau_1 d\tau_2
G_0(\tau_1-\tau_a)G_0(\tau_2-\tau_b)C(\tau_1,\tau_2), \label{corr}
\ee
where $G_0(\tau_2-\tau_1)=e^{-\omega_a|\tau_2-\tau_1|}/(2m\omega_0)$
and $C(\tau_1,\tau_2)= e^{-\Delta S(\tau_1,\tau_2)}$ with $\Delta S$
the action change due to the insertion of two instantons at $\tau_1$
and $\tau_2$. In the two different instanton (IST) phases
\begin{eqnarray}
\Delta S(\tau_1,\tau_2)=\left\{\begin{array}{cc}
c_1+c_2\ln|{\tau_2-\tau_1\over\tau_s}|
e^{-|{\tau_2-\tau_1\over\tau_s}|}&
\textrm{IST metal}\\
c_3\ln|{\tau_2-\tau_1\over\tau_0}|.&\textrm{IST insulator
}\end{array}\right.\nonumber\label{tp}
\end{eqnarray}
where $\tau_s$ is an instanton ``screening length'', and $c_1, c_2,
c_3$ are positive constants, implying
\begin{eqnarray}
C(\tau_1,\tau_2)\stackrel{|\tau_2-\tau_1|\rightarrow\infty}{\longrightarrow}
\left\{\begin{array}{cc} e^{-c_1} &
\textrm{IST metal}\\
0. & \textrm{IST insulator}\end{array}\right.\nonumber
\end{eqnarray}
Substituting the above asymptotic behavior of $C(\tau_1,\tau_2)$
into \Eq{corr} and noting the exponential decaying nature of
$G_0(\tau_1-\tau_2)$, we conclude that  the molecule has a
long/short range $\langle x(\tau)x(0)\rangle$ correlation function
in the instanton metal/insulator phase, respectively.

\begin{figure}
\includegraphics[angle=0,scale=0.52]{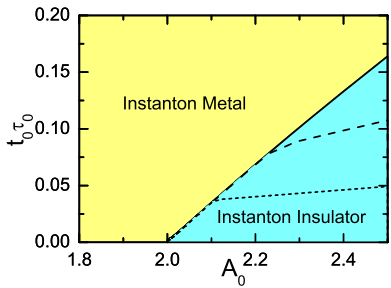}
\caption{(Color online) The zero temperature phase diagram in
$A_0-t_0\tau_0$ plane. In constructing it we fix the value of
$g_0/t_0$ to $0.3$. The short-dashed, dash, and solid lines are
the phase boundaries separating the instanton metal (upper left)
and instonton insulator (lower right) phases for
$\omega_0\tau_0=0.3$, 0.6, and $\omega_0\tau_0\ra \infty$,
respectively.\label{fig:epsart}}
\end{figure}

This suggests that in the instanton metal phase the molecule
essentially behaves as if it is in isolation (i.e., the electronic
eigenstates are eigenstates of $\s_z$ and the structure degrees of
freedom shows non-zero $\langle x\rangle$). In contrast, in the
instanton insulator phase the coupling between the electronic
state and the environment dominates. As a result, the electronic
eigenstates are essentially the eigenstates of $\s_x$ and the two
distinct molecular structures are averaged out. In the former case
the molecule retains its PI functionality, while in the latter
case the functionality is lost. This phase transition can be
realized by, e.g., tuning the strength of the molecule-environment
coupling ($A_0$ in Fig.~2.).

The transition between these two cases will be manifested as a
quantum phase transition at absolute zero temperature. This phase
transition can be detected via optical absorption spectrum. To
compute such spectrum we note that in the two different instanton
phases the spin-spin correlation function behaves
\begin{eqnarray}
\Gamma_{ss}(\tau)
\stackrel{|\tau|\rightarrow\infty}{\longrightarrow}
\left\{\begin{array}{cc} a_1 \Big({\tau\over\tau_s}\Big)^{-\alpha}
e^{-\tau/\tau_s} &
\textrm{IST metal}\\
a_2+a_3 \Big({\tau\over\tau_s}\Big)^{-\alpha} e^{-\tau/\tau_s} &
\textrm{IST insulator}\end{array}\right. \label{ss}\end{eqnarray}
Since $\Gamma_{ss}(\tau)$ is the Laplace transform of the optical
absorption spectrum, i.e.,$
\Gamma_{ss}(\tau)\equiv\int_0^\infty\sigma(\omega)e^{-\tau\omega}d\omega,
$
\Eq{ss} implies \be \sigma(\w)=a_2\delta(\w)+a_4\theta(\w-\Delta)
(\w-\Delta)^{\alpha-1}, \ee where $\Delta=1/\tau_s$, the
zero-frequency delta-function weight $a_2$ is nonzero only in the
instanton insulator phase, and $a_4$ is a constant proportional to
$a_1$ or $a_3$ depending on whether the system is in the instanton
metal or insulator phase. At the quantum critical point $\tau_s$
diverges and the optical absorption gap closes. As the molecule
loses the PI functionality a zero frequency peak should appear in
the optical absorption spectrum. These features can in principle
be detected optically. At non-zero temperatures, the quantum phase
transition will be smeared out. In the $A_0-T$ plane (assuming we
reach the phase transition by tuning $A_0$) there will be a region
in which $k_BT>1/\tau_s(T)$ where the molecule exhibit quantum
critical characteristics.

So far we have discussed the simplest model for PI. In reality
there could be more than one structural modes that couple to the
electronic optical transition\cite{tiago}. We have investigated a
two-oscillator model\cite{mckenzie,hahn} and reach the conclusion
that so long as the two structure oscillator are non-degenerate,
the results discussed above remain qualitatively unchanged.

In conclusion we have presented a theory of photoisomerization in an
environment. It points to the crucial role the environment plays in
this fascinating molecular quantum mechanics.


\acknowledgements CQW and JXL thank the support of the Berkeley
Scholar program and the NSF of China. CQW was also partially
supported by MOE of China (B06011). DHL was supported by the
Directior, Office of Science, Office of Basic Energy Sciences,
Materials Sciences and Engineering Division, of the U.S. Department
of Energy under Contract No. DE-AC02-05CH11231.


\begin{thebibliography}{99}
\bibitem{note}  It is often stated that there are
three electronic states responsible for two independent path ways
in the PI of azobenzene. From this point of view our work should
be interpreted as addressing one of the path ways.
\end{thebibliography}
\end{document}